\documentclass[journal=jctcce,manuscript=communication]{achemso}
\setkeys{acs}{maxauthors=0,articletitle=true}
\usepackage{achemso}
\setkeys{acs}{maxauthors=10, etalmode=truncate}
\usepackage{placeins}
\usepackage{graphics}
\usepackage{amssymb,amsfonts}
\usepackage{graphicx}
\usepackage[table,dvipsnames]{xcolor}
\usepackage{multirow}
\usepackage{caption}
\usepackage{subcaption}
\usepackage{booktabs}
\usepackage{colortbl}
\usepackage{amsmath}
\usepackage{amsopn}
\usepackage{bm}
\usepackage{braket}
\usepackage{siunitx}
\usepackage{color}
\usepackage{array}
\usepackage{lscape}
\usepackage{mciteplus}
\usepackage[version=3]{mhchem}
\usepackage{ulem}
\usepackage{listings}
\usepackage{enumerate}
\usepackage{lmodern}
\usepackage{tikz}
\usepackage{mhchem}
\usepackage[implicit=false]{hyperref}

\usetikzlibrary{backgrounds}

\author{Jonas Greiner}
\affiliation[mainz]{Department Chemie, Johannes Gutenberg-Universit{\"a}t Mainz\\Duesbergweg 10--14, 55128 Mainz, Germany}
\author{Janus J. Eriksen}
\email{janus@kemi.dtu.dk}
\affiliation[dtu]{DTU Chemistry, Technical University of Denmark\\Kemitorvet Bldg. 206, 2800 Kgs. Lyngby, Denmark}

\title[TITLE]{Symmetrization of Localized Molecular Orbitals}
\begin{document}

\begin{abstract}

We present a novel algorithm for (i) detecting approximate symmetries inherently present among spatially localized molecular orbitals and (ii) enforcing these in numerically exact manners by means of unitary optimization techniques. The vast potential of our algorithm to compress a full set of molecular orbitals into only a minimal set of symmetry-unique orbitals is demonstrated, starting from localized bases of either Pipek-Mezey or Foster-Boys orbitals. Comparisons of results based on either of these two localization procedures indicate that Foster-Boys molecular orbitals can be spanned by a smaller number of symmetry-unique orbitals on average, making these outstanding candidates for the exploitation of general, (non-)Abelian point-group symmetries in a range of local correlation methods. As an illustration of said compressibility, our algorithm is capable of identifying a mere 14 symmetry-unique orbitals for the buckminsterfullerene in the highly symmetric $I_h$ molecular point group, corresponding to only $1.7\%$ of its total 840 molecular orbitals in a standard double-$\zeta$ basis set. The present work thus marks an important advancement in the exploitation of point-group symmetry within local correlation methods, for which the appropriate adaption of symmetry uniqueness among orbitals has the potential to yield unprecedented speed-ups.

\end{abstract}

\newpage

\section{Introduction}\label{intro_sect}

The use of (unitary) transformations as a means to change from one conceptual frame to another is ubiquitous in quantum chemistry. A prime example is the change from Cartesian to spherical polar coordinates when solving the electronic Schr{\"o}dinger equation for the hydrogen atom, allowing for the efficient separation of variables. In modern chemical dynamics and as an important vehicle for the theoretical interpretation of molecular spectroscopy, the interchange from an adiabatic to any of a wealth of diabatic representations of electronic potential energy surfaces offers another example of particular practical value whenever the fundamental Born-Oppenheimer approximation becomes ill-defined. In both cases, alternative modes of representation will offer the same basic information, but ease of interpretation can generally vary. The same holds true for transformations between different molecular orbital (MO) bases in mean-field electronic-structure theories, made possible by the fact that Slater determinants stay invariant under a unitary rotation of the occupied MOs from a converged solution in Hartree-Fock (HF) or Kohn-Sham density functional theory (KS-DFT).\\

Although the two are equivalent in the sense that they give rise to the same mean-field properties, spatially localized MOs occasionally offer original modes of elucidation over conventional canonical orbitals. For instance, non-canonical MOs have been used to complement the conceptual depiction of bond rearrangements in organic chemistry by an arguably more rigorous basis in physical chemistry~\cite{Knizia2015}, and localized MOs have been transformative within local correlation schemes where the minimization of orbital spreads is integral for screening purposes~\cite{Saebo1993}. The more succinct spatial properties notwithstanding, a major drawback of spatially localized MOs is, however, concerned with a general inability to exploit molecular point-group symmetries in this representation. For both mean-field and correlation methods that instead operate in a basis of standard canonical orbitals, the construction of a set of symmetry-adapted orbitals may typically improve significantly upon the efficacy of computational simulations. These transform as irreducible representations of a given molecular point group, enabling integrals over totally symmetric operators to vanish whenever the direct product of irreducible representations of the corresponding orbitals is not totally symmetric. As a direct implication, entire blocks of integrals and cluster amplitudes can be made to vanish, whereby storage requirements and, in turn, the computational effort required to handle and contract these quantities can be significantly reduced~\cite{Pitzer1973,Stanton1991}. Spatially localized orbitals, on the other hand, are never constrained to transform as irreducible representations of molecular point groups as this would risk suppressing their overall spatial locality.\\

\begin{figure}[ht!]
    \centering
    \includegraphics[width=\textwidth]{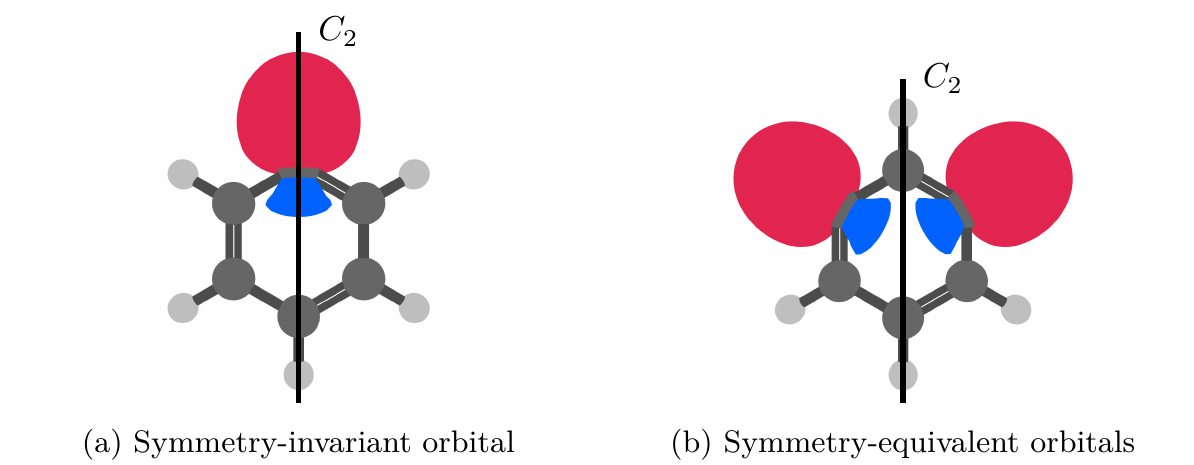}
    \caption{Examples of localized orbitals that are symmetric around a $C_2$ rotation axis.}
    \label{fig_1}
\end{figure}
Be that as it may, many localized MOs of symmetric molecules nevertheless do exhibit specific near-symmetric properties, and in the present study we will demonstrate how to exploit these (by enforcing exact rather than approximate symmetries, {\textit{vide infra}}) without penalizing spatial locality at all. As we will demonstrate, the present algorithm will be able to symmetrize MOs in both Abelian and higher-order point groups. Throughout, we will generally distinguish between so-called {\textit{symmetry-invariant}} and {\textit{symmetry-equivalent}} orbitals, cf. Fig. \ref{fig_1} for informative examples of what such symmetries may look like in the case of benzene. An orbital will be denoted as symmetry-invariant whenever it is localized on any point, axis, or plane of symmetry, as illustrated in the left panel of Fig. \ref{fig_1}. In cases where this is not true, MOs may instead be classified as equivalent with respect to specific symmetry operations of the molecular point group~\cite{LennardJones1949}. As is evident from the right panel of Fig. \ref{fig_1}, such orbitals do not transform into themselves under a given operation, but rather into other orbitals within a given considered subspace. Unfortunately, localized orbitals generally fail to be exactly symmetry-equivalent. Particularly for virtual orbitals of large one-electron basis sets, which are notoriously difficult to localize, symmetry properties will often differ from what would perhaps be expected. In many cases, orbitals exhibit only approximate symmetry properties due to the fact that cost functions in different localization schemes tend to have numerous, densely clustered local minima that are themselves not necessarily required to be symmetric~\cite{Subotnik2004,Subotnik2005}. While this might be of no particular relevance in some cases, say, if only the general topological features of a set of MOs are of interest, other applications will necessarily mandate strict exactness of symmetries, e.g., if certain numerical quantities (such as, integrals, coefficients, amplitudes, etc.) are to be assumed perfectly identical.\\

Of key importance in the context of the present work is the fact that a set of symmetry-equivalent MOs will require only a single orbital to construct all other members of a particular subspace through the application of a molecular point group's various symmetry operations. As such, every symmetry-equivalent set of MOs requires only a single {\textit{symmetry-unique}} orbital from which to span the set in its entirety. If implemented appropriately, this property may facilitate the equivalence of specific integrals and related quantities for many routine quantum-chemical methods whenever certain MOs can be transformed into one another using a befitting set of symmetry operations. Redundancies such as these can be computationally exploited using the so-called petite-list method by Dupuis and King~\cite{Dupuis1977}, and the savings afforded by such an algorithm will, in principle, be approximately inversely proportional to the full order of the point group. The actual number of symmetry-unique orbitals, however, will remain contingent on both molecular structure and the choice of localization procedure.\\

As alluded to above, localized MOs will generally fail to exhibit perfect symmetries on account of whatever exact cost objective function regulates their optimization. As a solution to this issue, we here devise a computational scheme that imposes {\textit{exact}} symmetry properties among a set of orbitals without disrupting the overall spatial locality of the set. Our proposed algorithm starts from an initial set of localized MOs, from which it attempts to find appropriate unitary transformations capable of constructing a set of orbitals that realizes any approximate symmetry properties inherently present in the original set of MOs. In a similar vein, K{\"o}ppl and Werner have previously proposed manners by which approximately symmetry-equivalent MOs can undergo a non-unitary averaging to ensure exact symmetry properties of localized orbitals~\cite{Koeppl2015}. Such transformations are generally sufficient for the symmetrization of occupied orbitals, given that deviations from exact symmetries tend to be small for these, but the non-unitary nature of the approach will formally break orthogonality. In the virtual space, on the other hand, the use of projected atomic basis functions will implicitly alleviate any deviations from symmetry. The preservation of symmetry during orbital localization has also been achieved by penalizing the absence of both spatial and energy locality of MOs during the optimization process~\cite{Su2020}, and this process will generally avoid the mixing of orbitals not related by symmetry. While placing additional constraints on a localization cost function with respect to energy locality will incur no additional benefits barring a reduction of the delocalization error in KS-DFT~\cite{Li2017}, such modifications will lead to less spatially local orbitals than those produced by standard localization cost functions.\\

What instead distinguishes our algorithm from these aforementioned methods is the fact that we are constructing localized, orthogonal mean-field orbitals in either of the occupied and virtual subspaces that satisfy exact symmetry properties. For the common choices of localization schemes assessed herein, we have found deviations from symmetry (in arbitrary point groups) following localization not to be negligible. For the purpose of using localized orbitals to reduce the scaling of correlation methods, the exactness of such symmetry properties must necessarily be of greater importance than how spatially confined these are, given how non-equivalences among supposedly symmetrical MOs will cause incorrect results up to a considered numerical precision. That being said, we here demonstrate how the locality of the MOs produced by our symmetrization procedure is not impacted and that these are as appropriate for use in local correlation methods as those of any parent localization scheme.

\section{Theory}\label{theory_sect}

In the present work, we seek to symmetrize a set of either occupied or virtual localized MOs by minimizing an appropriate objective function, $\mathcal{J}$, which depends on the approximate symmetry properties of an initial set of orbitals. Given some symmetry operation, $\hat{G}$, the corresponding integrals for the associated symmetry transformation, $\braket{\phi_p|\hat{G}|\phi_q}$, will describe the overlap between an initial orbital, $\phi_p$, and its transformed counterpart, $\hat{G}\phi_q$. Our objective function must be designed to impose a penalty whenever certain blocks of the matrix representation of $G_{pq}$ fail to vanish, while vanishing itself when all symmetry constraints are fulfilled. Phrased differently, in the case a set of orbitals, $S$, is either equivalent or invariant with respect to said symmetry operation, the matrix ($\bm{G}$) will have non-vanishing elements only for the set of orbitals that these orbitals are transformed into. By expressing rotations of an initial set of localized MOs, $\{\phi\}$, through the action of the exponential of an anti-Hermitian operator, $\kappa$, onto these~\cite{mest}, while further expanding all MOs into the constituent atomic orbitals (AOs) of our basis set, $\{\chi\}$, our objective function will read as follows
\begin{align}
\mathcal{J}=\sum_G^h\sum_{(S_1, S_2)\in \mathcal{S}_G}\sum_{p\in S_1}\sum_{\bar{q}\notin S_2}\big(\sum_{tu\mu\nu}\exp(\bm{\kappa}^\dagger)_{tp}\exp(\bm{\kappa})_{u\bar{q}}c_{\mu p}^*c_{\nu\bar{q}}\braket{\chi_\mu|\hat{G}|\chi_\nu}\big)^2 \ . \label{obj_func}
\end{align}
In Eq. \ref{obj_func}, the set of tuples, $\mathcal{S}_G$, includes all orbitals, $S_1$, that transform into another orbital set, $S_2$, for a symmetry operation, $\hat{G}$, and the power of 2 is implemented so as to avoid the cancellation between contributions to $\mathcal{J}$ from different groups of orbitals or symmetry operations. If instead an absolute function had been implemented for this purpose, then problems of exaggerated step sizes could potentially arise in the vicinity of a minimum since the modified cost function would yield a large gradient while remaining non-differentiable. The cost function itself is minimized by means of unitary optimization~\cite{Lehtola2013}, and the employed parametrization guarantees full orthonormality among the orbitals of the transformed set.

\section{Algorithm}\label{algo_sect}

Our proposed objective function cannot be solved on the basis of gradient information alone using a root-finding algorithm since the desired root has a multiplicity of 2 which will cause instabilities. Instead, Eq. \ref{obj_func} is minimized using the co-iterative augmented Hessian method (CIAH)~\cite{Sun2016}. The gradient, the diagonal of the Hessian, as well as the contraction between the Hessian and some trial function are all included in the supporting information (SI). To ensure that the orbitals can be symmetrized close to the limit of the floating point representation ($10^{-14}$), the cost function, gradient, and Hessian have all been scaled by a factor of $\sqrt{10^{14}}$.\\

The minimization of Eq. \ref{obj_func} risks being excessively expensive for large systems, sizable basis sets, and higher point groups as the number of symmetric orbital sets in $\mathcal{S}_G$ and/or the order of the point group, $h$, increase. Further to that, calculating the gradient and Hessian of Eq. \ref{obj_func} requires access to and multiplication of elements of the $\bm{G}$ matrix that stride out of continuous memory order. For this reason, we propose the following two-step algorithm: First, orbitals are divided into sets that are approximately symmetry-invariant with respect to all symmetry operations, and the orbitals are next fully symmetrized with respect to these symmetry constraints. Since $\mathcal{S}_G$ is the same for all symmetry operations, orbitals can be reordered and accessed in slices, which significantly improves performance. In a second step, the orbital space spanned by the separate sets then gets symmetrized with respect to the individual symmetry operations. These sets are bound to be significantly smaller than the full orbital space, leading to an acceleration of the overall symmetrization.\\

\begin{figure}[ht!]
    \includegraphics[width=\textwidth]{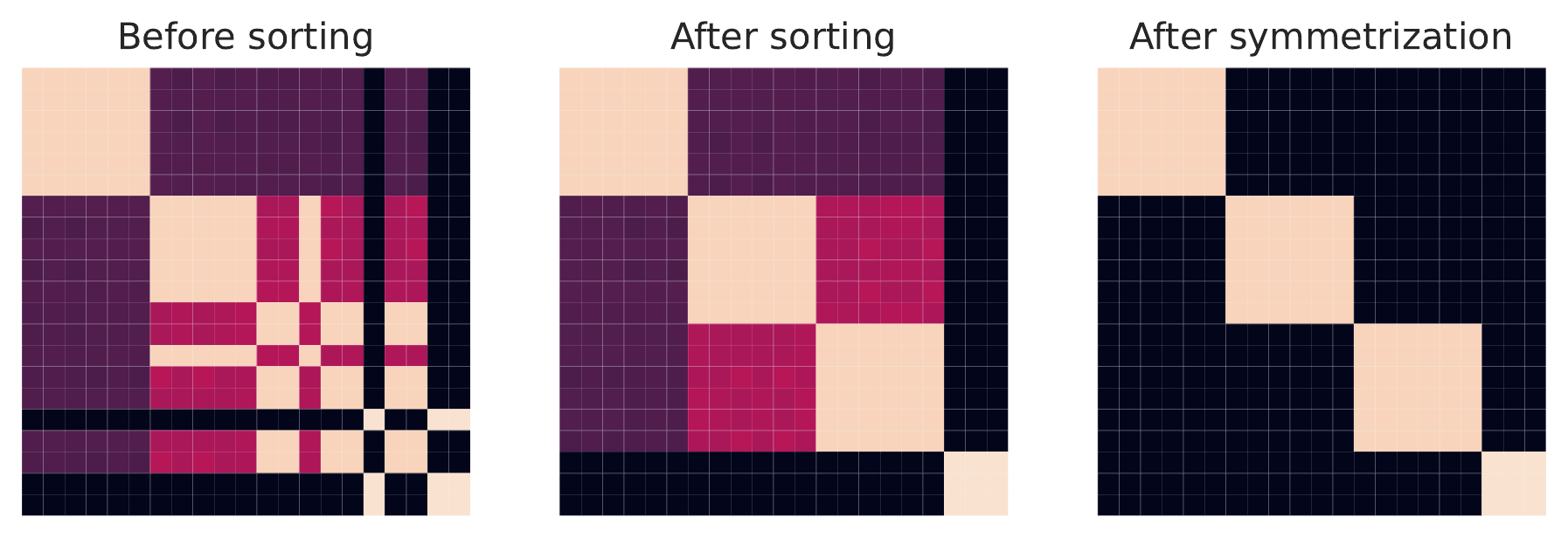}
    \caption{Heatmaps of ${\bm{G}}^{\Sigma}$ for Pipek-Mezey localized occupied orbitals of benzene before symmetrization, after sorting according to the reverse Cuthill-McKee algorithm, and after symmetrization. Orbital indices are plotted along both the x-/y-axes and the employed color map is logarithmic, ranging in values of ${\bm{G}}^{\Sigma}$ from $10^{-13}$ (black) to $1$ (bright beige tone).}
    \label{fig_2}
\end{figure}

Approximate symmetries present in a set of orbitals can be detected by locating elements that nearly vanish in the overlap matrix between initial and transformed orbitals, $G_{pq}$. In the first step of our symmetrization procedure, the absolute value of $\bm{G}$ is summed for all symmetry operations to produce a new matrix quantity, ${\bm{G}}^{\Sigma}=\sum_G|\bm{G}|$. Independent blocks of this matrix will be invariant with respect to all symmetry operations, and its block structure can be revealed using the reverse Cuthill-McKee algorithm~\cite{Cuthill1969, George1981}, which yields the orbital order necessary to form a block-diagonal matrix with a sufficiently small bandwidth. In our work, its implementation in the {\texttt{SciPy}} library is used with a certain cut-off parameter~\cite{Virtanen2020}
\begin{align}
\lambda_{p}=\gamma\max_qG_{pq} \ .
\end{align}
The input parameter, $\gamma$, describes the ratio of the maximum orbital contribution below which an orbital, $\phi_p$, is not considered to contribute to a transformed orbital, $\phi_p$. Below this threshold, blocks of ${\bm{G}}^{\Sigma}$ are regarded as zero for the purpose of the symmetry detection algorithm, and an appropriate tuning of this parameter can thus influence the degree to which approximate symmetries need be present in an initial set of localized orbitals for the algorithm to consider them. Next, Eq. \ref{obj_func} can then be used to find a unitary rotation which, when applied to these orbitals, renders all of those contributions to ${\bm{G}}^{\Sigma}$ falling below $\lambda_{p}$ as insignificant. As a result, all individual orbital spaces that were only approximately symmetry-invariant with respect to all symmetry operations to begin with end up invariant up to the convergence criterion of the CIAH algorithm. As an illustration of our procedure, Fig. \ref{fig_2} shows values of ${\bm{G}}^{\Sigma}$ for a set of localized, occupied orbitals of benzene before symmetrization, after sorting of the orbitals using the reverse Cuthill-McKee algorithm, and after the final symmetrization.\\

The symmetry properties for the second step of our algorithm can be detected with a similar, albeit somewhat more elaborate algorithm. For every symmetry operation and every orbital, orbital contributions above the cut-off parameter, $\lambda_{p}$, are recorded. This record describes the set of orbitals an orbital $p$ transforms into when a certain symmetry operation is applied to it. As an example, this will amount to just a single orbital in the case of symmetry-equivalent orbitals, cf. Fig. \ref{fig_1}. For cases where a set of multiple orbitals transforms into another set, lists of orbitals for every single symmetry operation need to be compared to find intersections between sets of transformed orbitals, and the involved sets of orbitals get combined for each of these. After considering the overlap of all combinations of orbitals, multiple tuples of two orbital sets $(S_1,S_2)$ are recovered, describing sets of orbitals ($S_2$) that the orbitals of $S_1$ approximately transform into under the symmetry operation. Finally, the exactness of these symmetries can again be enforced by minimizing Eq. \ref{obj_func}.

\section{Computational Details}\label{comp_sect}

In the following, all analyses of our symmetrization procedure are based on a selection of molecular systems of different sizes that belong to a variety of point groups: ammonia (\ce{NH3}, $C_{3v}$), methane (\ce{CH4}, $T_d$), boron trifluoride (\ce{BF3}, $C_{3h}$), bromine pentafluoride (\ce{BrF5}, $C_{4v}$), sulfur hexafluoride (\ce{SF6}, $O_h$), benzene (\ce{C6H6}, $D_{6h})$, glucose (\ce{C6H12}, $D_{3d}$), pentaerythrityl tetrachloride (\ce{C5H8Cl4}, $S_4$), sulfur (\ce{S8}, $D_{4d}$), and buckminsterfullerene (\ce{C60}, $I_h$). All results are based on closed-shell HF orbitals using the cc-pVDZ basis set~\cite{dunning1989}, and the localization schemes used throughout are Pipek-Mezey~\cite{Pipek1989} (PM), Foster-Boys~\cite{Foster1960} (FB), and Edminston-Ruedenberg~\cite{Edmiston1963} (ER), with orbitals optimized according to the defaults of the {\texttt{PySCF}} program~\cite{Sun2017, Sun2020}. In all orbital localizations, an initial guess is spanned by a set of converged HF orbitals described through the closest unitary transformation to the orthogonalized AO basis in a least-squares sense~\cite{Ziolkowski2009, Sun2016}. The code used to perform the actual symmetrization is our newly-developed, open-source {\texttt{SymLo}} Python code~\cite{SymLo}, which further leverages existing functionalities of {\texttt{PySCF}} for second-order optimizations, the transformation of spherical AOs using Wigner D matrices, and the general handling of molecular point-group symmetry.\\ 

Our symmetrization of a set of localized orbitals is assumed converged whenever the maximum absolute element of ${\bm{G}}$ contributing to the cost function in Eq. \ref{obj_func} falls below a predefined threshold. A convergence criterion of $u_{\text{block}} = 10^{-13}$ is used for the block symmetrization and $u_{\text{orb}} = 10^{-12}$ is used for the subsequent symmetrization steps on individual orbital spaces. For both of these steps, a dedicated parameter governs the symmetry detection ($\gamma_{\text{block}}$ and $\gamma_{\text{orb}}$) to permit for greater direct control~\bibnote{For both PM and FB orbitals, $\gamma_{\text{block}} = \gamma_{\text{orb}} = 0.3$ for all systems, except for \ce{BF3} and \ce{SF6} in the case of FB, for which required parameter sets of $\gamma_{\text{block}} = 0.4/\gamma_{\text{orb}} = 0.1$ and $\gamma_{\text{block}} = 0.3/\gamma_{\text{orb}} = 0.2$ were needed to converge the symmetrization, respectively}. As will be discussed in the following, the orbitals of the ER scheme generally fail to be symmetrized due to a pronounced lack of inherent symmetries present amongst these. Heatmaps of ${\bm{G}}^{\Sigma}$ are supplied in the SI for all systems based on either PM (Figs. S1 to S6) or FB orbitals (Figs. S7 to S12).\\

To give quantitative assessments of the spatial locality of an MO $p$, we will compare results in different representations using both second- and fourth-moment orbital spreads~\cite{Burke1978, Jansik2011, Hoeyvik2012}
\begin{align}
\sigma_n^p=\sqrt[n]{\Braket{p|\left(\bm{r}-\Braket{p|\bm{r}|p}\right)^n|p}}\qquad (n=2,4). \label{mom_eq}
\end{align}

\section{Results}\label{res_sect}

\begin{figure}[ht!]
    \includegraphics[width=\textwidth, trim={2.5cm 2.5cm 2.5cm 2.5cm}, clip]{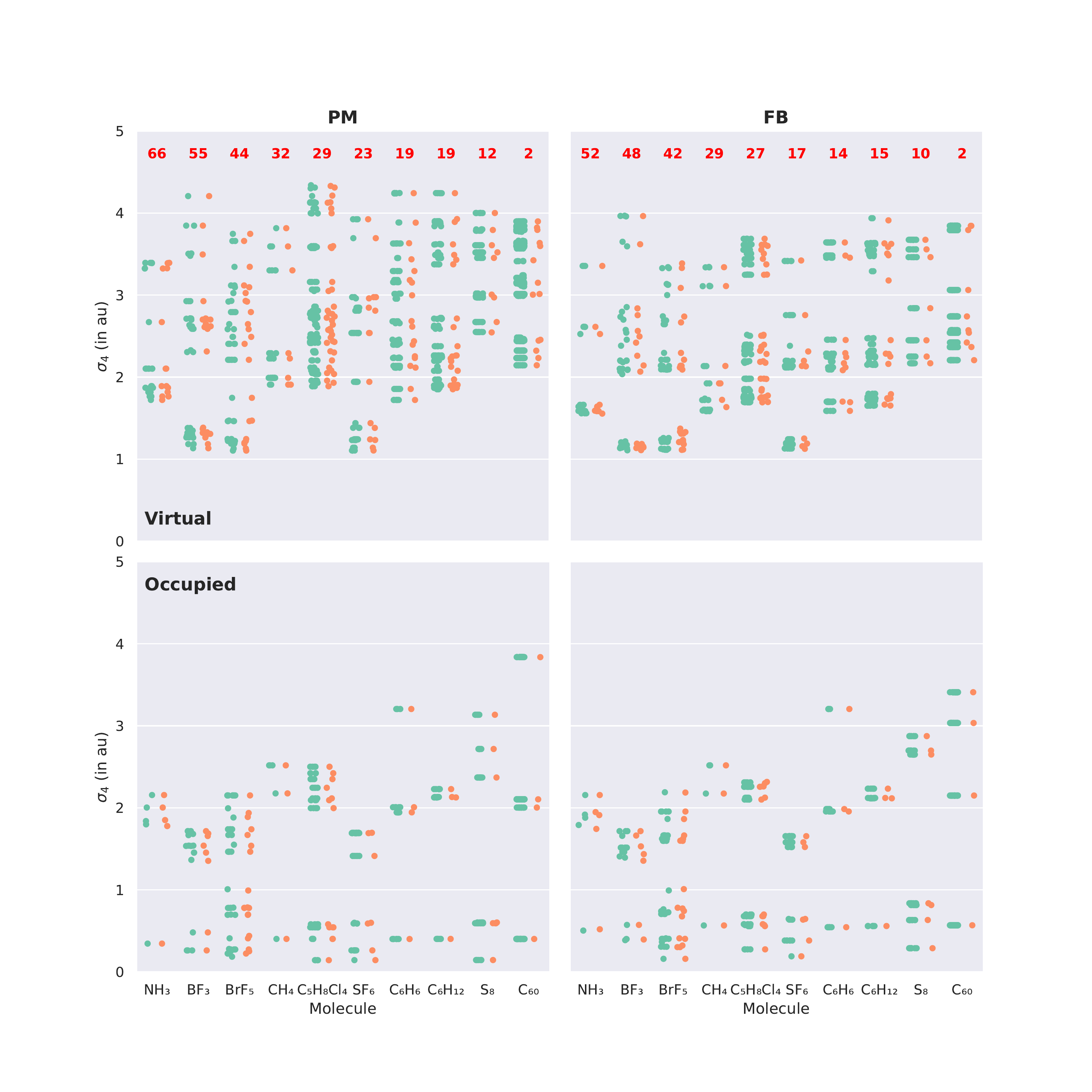}
    \caption{Fourth-moment orbital spreads ($\sigma_4^p$) of PM and FB localized orbitals. Results for orbitals before and after the symmetrization are displayed in green and orange, respectively, and the percentage of symmetry-unique orbitals for every molecule is displayed in red.}
    \label{fig_3}
\end{figure}
Given that our symmetrization procedure will yield a set of orbitals that do not necessarily correspond to a stationary point of a given native localization cost function, it is essential to any subsequent applicability within local correlation methods that spatial locality is not significantly affected by the applied rotations. In Fig. \ref{fig_3}, fourth-moment orbital spreads (Eq. \ref{mom_eq}) are compared for PM and FB localized orbitals both prior to and following symmetrization. (A similar plot for second-moment orbital spreads can be found in Fig. S13 of the SI). The comparisons in Figs. \ref{fig_3} and S13 leave no doubt as to whether or not locality is conserved since the bulks and tails of both sets of MOs for all considered systems remain virtually unaffected by the symmetrization. As is to be expected, virtual MOs are less local by both measures, but most noteworthy is perhaps the observation that different localization procedures will produce orbitals that satisfy different symmetry properties and differ in their spatial locality.\\

\begin{figure}[ht!]
    \includegraphics[width=\textwidth]{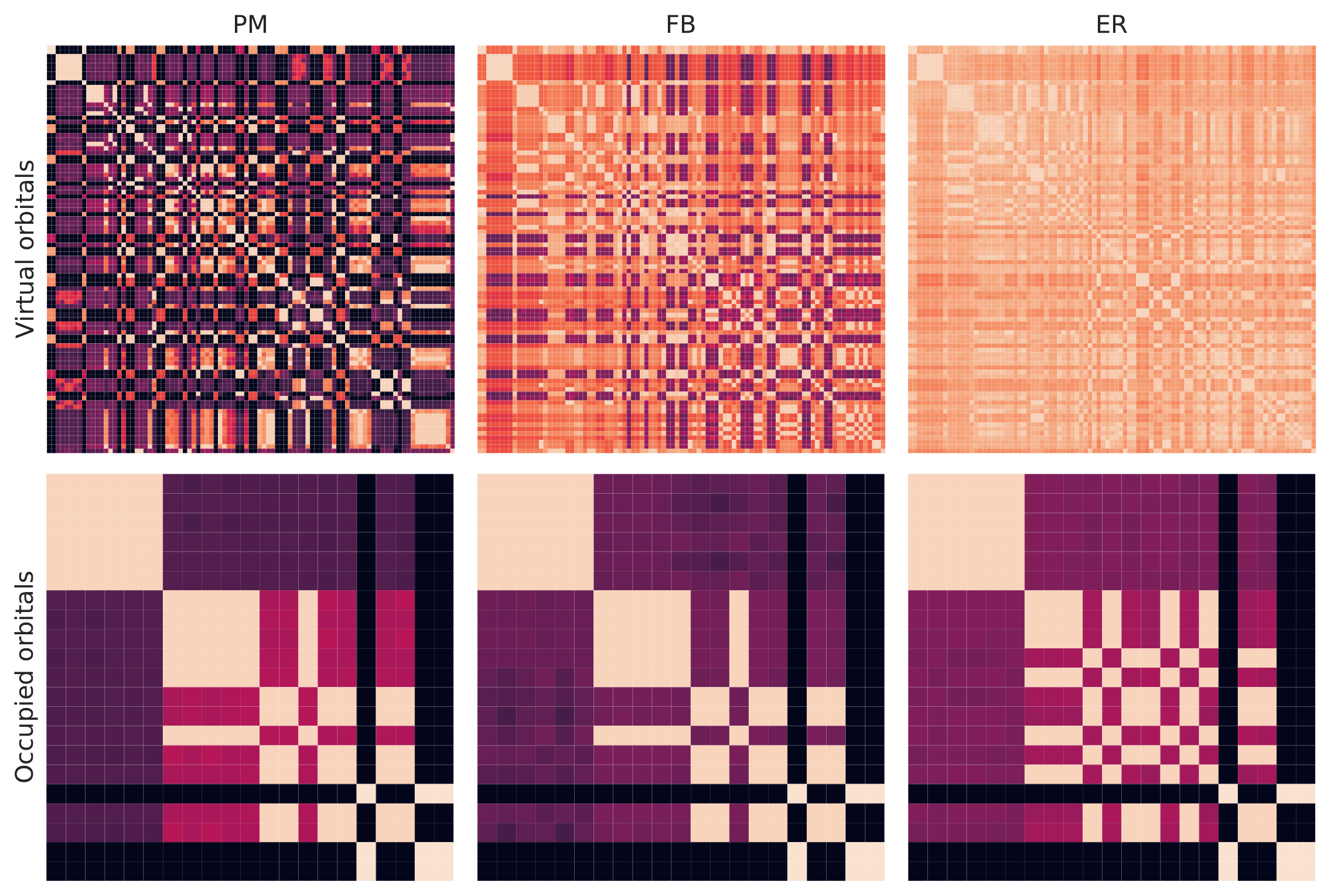}
    \caption{Heatmaps of ${\bm{G}}^{\Sigma}$ for different localized orbitals of benzene prior to symmetrization. Orbital indices are plotted along both the x-/y-axes and the color map is that of Fig. \ref{fig_2}.}
    \label{fig_4}
\end{figure}
In general, the symmetry detection for FB and, in particular, ER virtual orbitals is more difficult than for corresponding PM virtuals. For the localized orbitals of benzene in Fig. \ref{fig_4}, the initial localized orbitals produced by any of these two procedures tend to deviate from symmetry to a greater extent than what we observe in the case of PM. Such issues are at risk of making our symmetry detection more prone to errors and may even lead to convergence issues in the subsequent symmetrization step. While all symmetrizations of FB virtuals could be made to converge, this was not true for the ER virtual MOs of most of the systems considered herein. This is a consequence of the design of our algorithm and its formulation around symmetries which must already be inherently present among a set of MOs following their spatial localization. As such, the general inability of our algorithm to symmetrize ER localized orbitals is not problematic {\textit{per se}}, since the ER scheme itself can be deemed inferior to PM and FB in terms of computational scaling and the overall spatial locality of its MOs~\cite{Folkestad2022}.\\

The comparison of the symmetrization of PM and FB orbitals in Fig. \ref{fig_3} also warrants a closer look at the spatial locality of these as well as the total number of symmetry-unique orbitals required to span them. In terms of both second- and fourth-moment orbital spreads, the PM and FB localization procedures appear to produce orbitals that are very much on par, with a comparable degree of locality observed between the two. The percentage of symmetry-unique orbitals with respect to the whole set of MOs is also displayed in Fig. \ref{fig_3}. For all systems, the FB procedure yields a smaller overall ratio. Recalling that these MOs are based on the minimization of the second-moment orbital spread in Eq. \ref{mom_eq}, a possible explanation for this observation might be the absence of $\sigma$-$\pi$ separation among FB MOs which, in turn, leads to a set of so-called (symmetry-equivalent) $\tau$ orbitals.~\cite{Pipek1989}. Overall, the results in Fig. \ref{fig_3} seem to indicate that whenever spatial locality and orbital symmetry rule any kind of chemical interpretability, the FB scheme should be considered a strong alternative to the prevalent and widespread use of PM and its family of derivative localization schemes~\cite{lehtola2014}.\\

\begin{figure}[ht!]
    \includegraphics[width=0.65\textwidth]{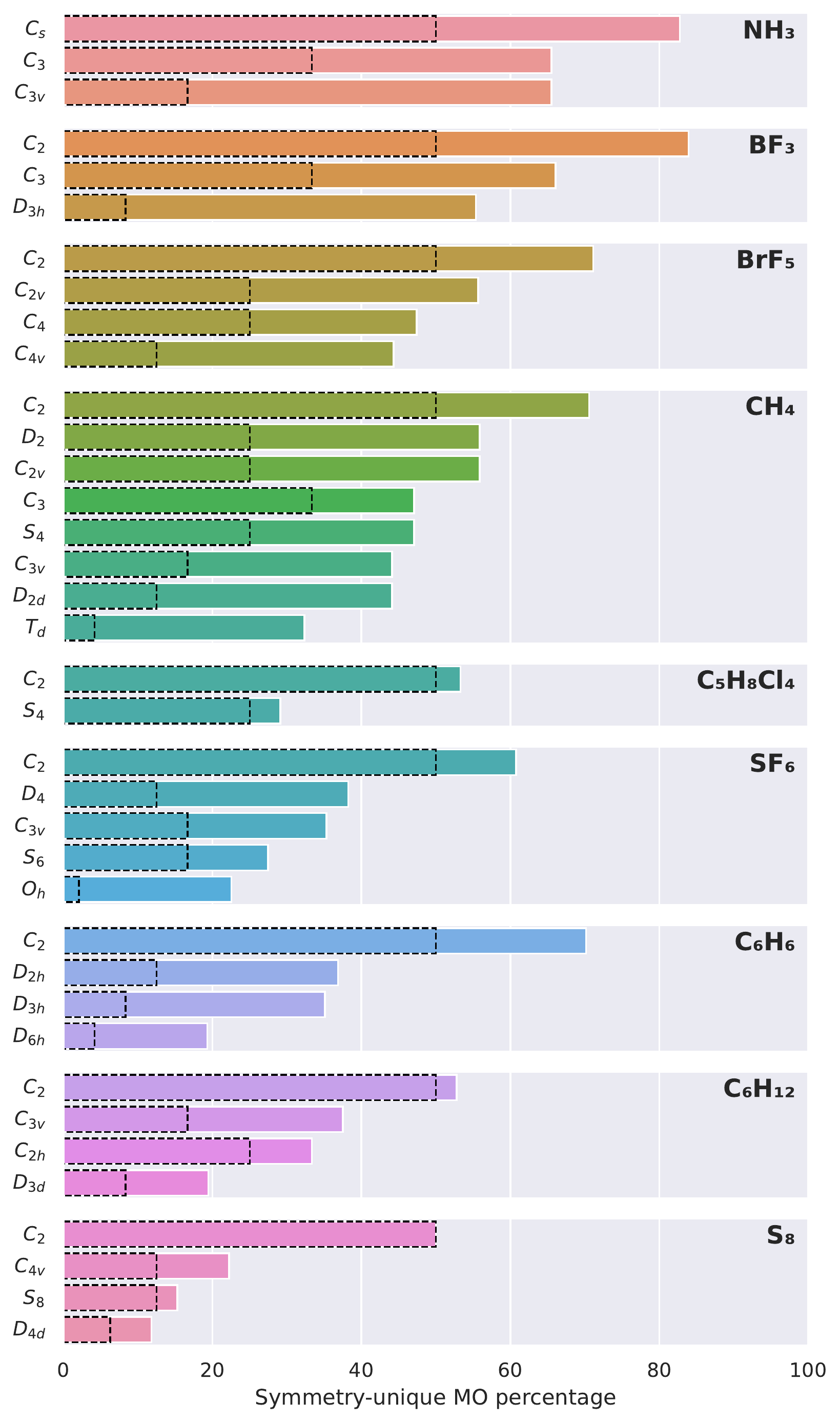}
    \caption{Ratio of symmetry-unique PM orbitals in the highest point group and selected subgroups for a range of molecules. Theoretical limits are superimposed using dashed lines.}
    \label{fig_5}
\end{figure}
Next, to illustrate the variance of results with respect to the point group of choice, symmetrizations were further completed for subgroups of the highest molecular point group to compare the resulting number of symmetry-unique orbitals, cf. Fig. \ref{fig_5}. Satisfactorily, the use of higher point groups produces a reduced number of symmetry-unique orbitals, and direct comparisons of the total number of these reveal some insight into the extent to which symmetry can be exploited in relation to the theoretical limit. As mentioned in Sect. \ref{intro_sect}, the reduction in computational time is inversely proportional to the order of the point group when exploiting symmetry in local correlation methods~\cite{Dupuis1977}. Given how all localization schemes considered herein seek to confine orbitals around atoms and bonds, the number of symmetry-unique orbitals is often also related to the number of symmetry-unique atoms in a molecular system at hand. Symmetry operations that fail to produce additional atoms when applied to non-redundant atoms in a given system will ultimately lead to a smaller number of symmetry-unique orbitals on this account, in comparison with the theoretical lower bound. This also explains why reductions are generally more moderate whenever atoms coincide with a point, axis, or plane of symmetry, an observation which is most evident when comparing the results in Fig. \ref{fig_5} for \ce{NH3} and \ce{C6H12} (in $C_{3v}$) or \ce{BrF5} and \ce{S8} (in $C_{4v}$).\\

\begin{figure}[ht!]
    \centering
    \includegraphics[width=\textwidth]{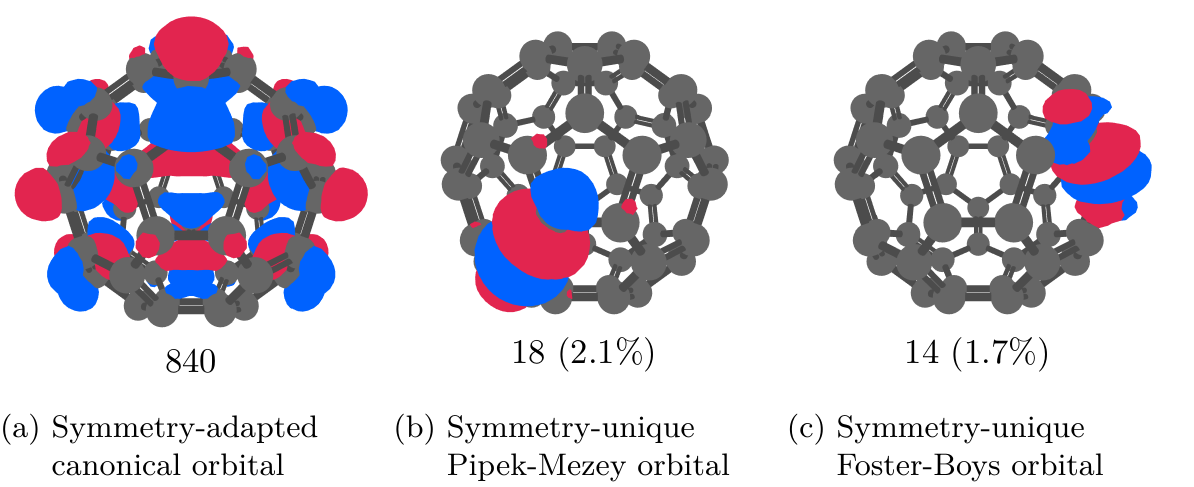}
    \caption{Molecular orbitals of buckminsterfullerene. The total number of symmetry-unique orbitals is displayed below each example illustration alongside relative ratios.}
    \label{fig_6}
\end{figure}
Finally, as an illustrative example of the power of our proposed symmetrization procedure, but also as a testament to its computational efficiency (cf. Sect. \ref{algo_sect}), we will end by complementing the results of Figs. \ref{fig_3} and \ref{fig_5} by also reporting numbers for the compression of the total space of MOs in the case of \ce{C60} in a cc-pVDZ basis set, which comprises a total of 840 orbitals. Buckminsterfullerene is the quintessential example of the highly symmetric $I_h$ molecular point group and, important in the present context, no atom is located at the center of symmetry, cf. the discussion around Fig. \ref{fig_5}. An appropriate example of its symmetry-adapted canonical orbitals is presented in Fig. \ref{fig_6}, as yielded by a standard HF calculation, in addition to symmetry-unique orbitals based on the PM and FB localization schemes. When starting from a set of localized MOs generated by either of these, a mere 18 or 14 symmetry-unique orbitals are needed to span the complete orbital space of orbitals, corresponding to a compression of nearly two orders of magnitude and close to the theoretical limit of $0.8\%$. Importantly, we once again find no sacrifices to spatial locality from our symmetrization, with mean differences in the second- and fourth-moment spreads both below $10^{-3}$ au, irrespective of the choice of underlying localization scheme, cf. Figs. \ref{fig_3} and S13. Collectively, these numbers indicate how the development of ingenious new local correlation schemes holds immense potential {\textit{if}} these are carefully designed in such manners that combinations of localized orbitals and point-group symmetries can be efficiently exploited.

\section{Discussion and Conclusions}

In the course of the present work, we have outlined a novel algorithm for the symmetrization of localized molecular orbitals and numerically demonstrated its applicability for both Pipek-Mezey and Foster-Boys orbitals. Our algorithm is able to detect symmetries inherently present in molecular orbitals upon localization, which it then enforces exactly by means of unitary optimizations. Comparisons of results based on initial sets of Pipek-Mezey and Foster-Boys orbitals strongly indicate how the latter tend to produce a smaller number of symmetry-unique orbitals from which to span a full set of molecular orbitals, making these excellent candidates for the exploitation of (non-)Abelian point-group symmetries in a range of local correlation methods~\cite{eriksen2021a}. These advancements thus promise to deliver unprecedented reductions to both storage and computational requirements by allowing befitting methods to benefit from an increased degree of sparsity in many common tensor quantities whenever these are expressed in a localized and appropriately symmetrized orbital basis. To that end, we are currently working on new ways in which to exploit arbitrary point-group symmetries using the petite-list method in the context of many-body expanded full configuration interaction theory~\cite{eriksen2021}. However, while symmetrized MOs will be affected by many of the same potential issues with topology (that is, spatial profile and extension) as their parent localized MOs~\cite{Hansen2020,Hoeyvik2020}, the symmetrization procedure of the present study will arguably also make the acceleration of many alternative local correlation methods feasible as long as these can be suitably adapted to leverage the exact orbital symmetries exposed by our new algorithm.

\section*{Acknowledgments}

This work was supported by a generous research grant (no. 37411) from VILLUM FONDEN (a part of THE VELUX FOUNDATIONS). The authors are grateful to Prof. J{\"u}rgen Gauss (JGU Mainz) for valuable comments and encouragement.

\section*{Supporting Information}

The supporting information (SI) collects formal expressions for the gradient of Eq. \ref{obj_func}, the diagonal of the Hessian, as well as the contraction between the Hessian and some trial function. Furthermore, the SI presents heatmaps of ${\bm{G}}^{\Sigma}$ for all systems but \ce{C60} based on either PM orbitals (Figs. S1 to S6) or FB orbitals (Figs. S7 to S12). Finally, Fig. S13 presents second-moment orbital spreads, on par with the fourth-moment spreads in Fig. \ref{fig_3}.

\section*{Data Availability}

Data in support of the findings of this study are available within the article and its SI.

\newpage

\providecommand{\latin}[1]{#1}
\makeatletter
\providecommand{\doi}
  {\begingroup\let\do\@makeother\dospecials
  \catcode`\{=1 \catcode`\}=2 \doi@aux}
\providecommand{\doi@aux}[1]{\endgroup\texttt{#1}}
\makeatother
\providecommand*\mcitethebibliography{\thebibliography}
\csname @ifundefined\endcsname{endmcitethebibliography}
  {\let\endmcitethebibliography\endthebibliography}{}


\begin{mcitethebibliography}{36}
\bibitem[Not()]{Note-1}
For both PM and FB orbitals, $\gamma_{\text{block}} = \gamma_{\text{orb}} =
  0.3$ for all systems, except for \ce{BF3} and \ce{SF6} in the case of FB, for
  which required parameter sets of $\gamma_{\text{block}} =
  0.4/\gamma_{\text{orb}} = 0.1$ and $\gamma_{\text{block}} =
  0.3/\gamma_{\text{orb}} = 0.2$ were needed to converge the symmetrization,
  respectively\relax
\mciteBstWouldAddEndPuncttrue
\mciteSetBstMidEndSepPunct{\mcitedefaultmidpunct}
{\mcitedefaultendpunct}{\mcitedefaultseppunct}\relax
\EndOfBibitem
\end{mcitethebibliography}
\end{document}